\documentclass{article}
\usepackage{spconf,amsmath,graphicx}

\usepackage[utf8]{inputenc} 
\usepackage[T1]{fontenc}    
\usepackage{hyperref}       
\usepackage{url}            
\usepackage{booktabs}       
\usepackage{amsfonts}       
\usepackage{nicefrac}       
\usepackage{microtype}      
\usepackage{xcolor}         

\usepackage{makecell}
\usepackage{adjustbox}
\usepackage{multirow}
\usepackage{mathtools}
\usepackage{amsmath}
\usepackage{caption}
\usepackage{subcaption}
\usepackage{lipsum}
\usepackage{wrapfig}
\usepackage{graphicx}
\usepackage[T1]{fontenc}

\usepackage[noabbrev,capitalize,nameinlink]{cleveref}
\usepackage[symbol]{footmisc}

\usepackage{xspace}

\makeatletter
\DeclareRobustCommand\onedot{\futurelet\@let@token\@onedot}
\def\@onedot{\ifx\@let@token.\else.\null\fi\xspace}

\def\ie{\emph{i.e}\onedot}

\def\etal{\emph{et al}\onedot}
\makeatother
\title{Perspective Projection-Based 3D CT Reconstruction from Biplanar X-rays}
\name{Daeun Kyung$^{1,\ast}$ \qquad Kyungmin Jo$^{1, \ast}$ \qquad Jaegul Choo$^1$ \qquad Joonseok Lee$^{2,3}$ \qquad Edward Choi$^{1, \dagger}$ \thanks{$^{\ast}$ The first two authors contributed equally.} \thanks{$^{\dagger}$ Corresponding author} \thanks{This work was supported by Institute of Information \& communications Technology Planning \& Evaluation (IITP) grant (No.2019-0-00075, No.2021-0-02068), Korea Health Industry Development Institute (KHIDI) grant (No.HI21C1138), Korea Medical Device Development Fund grant (Project Number: 1711138160, KMDF\_PR\_20200901\_0097), and National Research Foundation of Korea (NRF) grant (NRF-2021H1D3A2A03038607) funded by the Korea government(MSIT, MOTIE, MOHW, MFDS).}}
\address{
  $^1$KAIST \qquad $^2$Seoul National University \qquad $^3$Google Research
}

\begin{document}
\ninept
\maketitle
\begin{abstract}
X-ray computed tomography (CT) is one of the most common imaging techniques used to diagnose various diseases in the medical field. Its high contrast sensitivity and spatial resolution allow the physician to observe details of body parts such as bones, soft tissue, blood vessels, etc. As it involves potentially harmful radiation exposure to patients and surgeons, however, reconstructing 3D CT volume from perpendicular 2D X-ray images is considered a promising alternative, thanks to its lower radiation risk and better accessibility. This is highly challenging though, since it requires reconstruction of 3D anatomical information from 2D images with limited views, where all the information is overlapped. In this paper, we propose PerX2CT, a novel CT reconstruction framework from X-ray that reflects the perspective projection scheme. Our proposed method provides a different combination of features for each coordinate which implicitly allows the model to obtain information about the 3D location. We reveal the potential to reconstruct the selected part of CT with high resolution by properly using the coordinate-wise local and global features. Our approach shows potential for use in clinical applications with low computational complexity and fast inference time, demonstrating superior performance than baselines in multiple evaluation metrics.
\end{abstract}

\begin{keywords}
X-ray computed tomography, CT reconstruction
\end{keywords}
\section{Introduction}
X-ray computed tomography (CT) is a medical imaging technique that produces cross-sectional images of the body from multi-view X-ray projection data scanned around the patient. Due to its advantages of having high spatial and density resolution, it is widely used in the medical domain. Specifically, CT is helpful to diagnose diseases as it simultaneously shows details of body parts such as bones, soft tissue. However, CT scans have the disadvantage of incurring more radiation exposure than other medical imaging techniques~\cite{david2007ctrisk,power16ctrisk,Schmidt12ctrisk}. 

Previous work made effort to reduce the radiation dose by reducing the number of CT projections while maintaining the high quality of reconstruction based on traditional methods~\cite{Kudo13traditional} or deep learning approaches~\cite{Ye2018fewshotct, Wu21drone, zhang18sparsect, 14ctrecon}. However, these studies are limited to only marginally reducing the radiation dose, still requiring hundreds of projection images obtained from CT scanners to ensure high quality of images.
In addition, they still need a CT scanner which is less accessible than X-ray machines. Efforts to resolve these problems have led to attempts to reconstruct CT using X-rays obtained from traditional X-ray machines~\cite{Abril22mednerf, ge2022x, henzler17singlexray, jiang21cvae, kasten2020end, ratul2021ccx,shen19patrecon, ying2019x2ct}.

X-ray imaging is the most frequently performed radiographic examination in hospitals, as it needs less physical and economic burden on the patient. In particular, X-rays have less radiation exposure than CT and do not require additional preparation such as contrast agents. While CT requires the patient to lie on a cylindrical device, X-rays do not have these limitations, making them easier to use for patients with reduced mobility. Thus, it will be desirable to reconstruct 3D internal body information from X-rays.

The primary challenge of reconstructing the CT volume from the X-ray is the lack of depth information; estimating 3D structures from 2D data is a well-known ill-posed problem~\cite{bertero1988ill}. The 3D reconstruction problem gets more ambiguous as we use fewer X-rays, making it harder to solve with traditional CT reconstruction approaches. Recently, large-scale data and deep learning models~\cite{Abril22mednerf, ge2022x, kasten2020end, ratul2021ccx, ying2019x2ct} powered to tackle this ultra-sparse view reconstruction problem by learning prior knowledge of human anatomy.

\begin{figure}[t!]
\begin{center}
\includegraphics[width=0.9\linewidth]{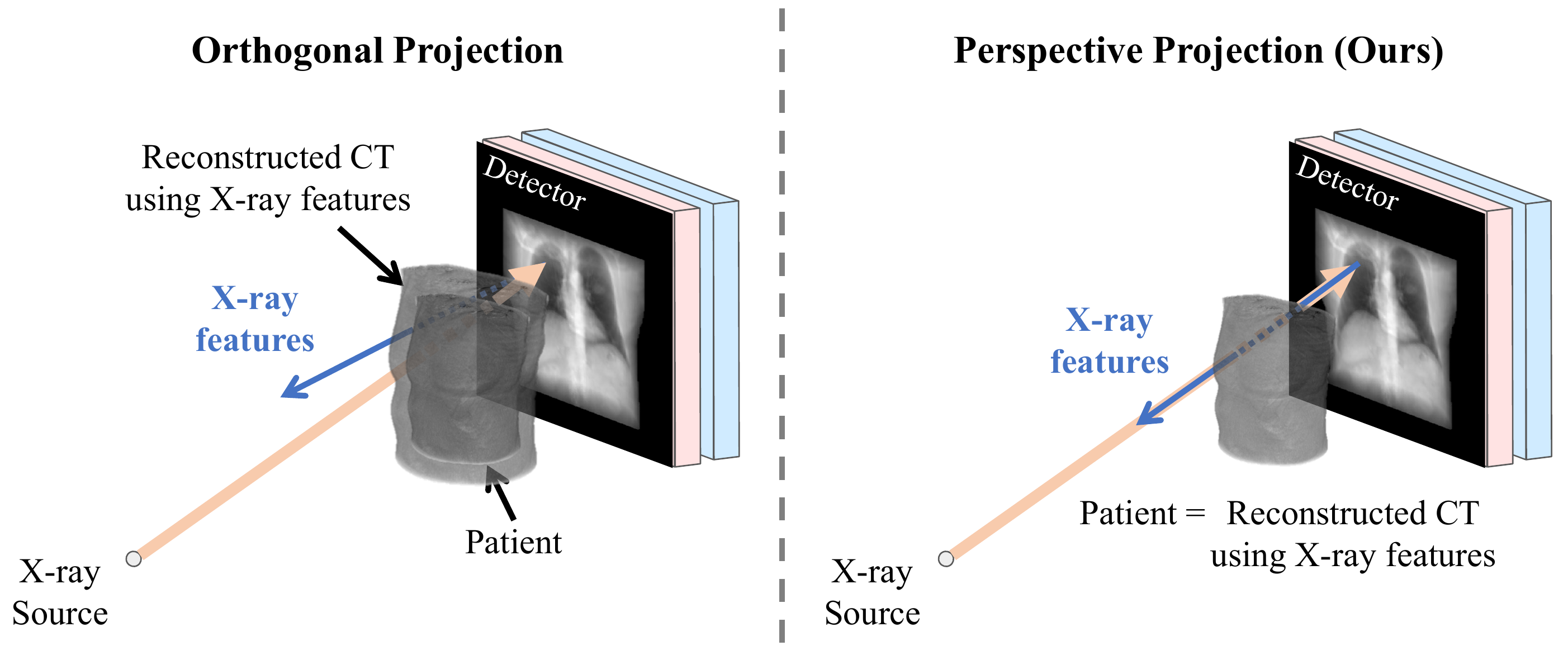}
\end{center}
\vspace{-0.4cm}
\caption{
(left) Orthogonal projection-based feature expansion. 
(right) Perspective projection-based feature extension. 
}
\vspace{-0.45cm}
\label{fig:projection}
\end{figure}

Some studies use a 2D-to-3D structure consisting of a 2D encoder and a 3D decoder to provide depth information from a few X-ray images~\cite{ge2022x, kasten2020end, ratul2021ccx, ying2019x2ct}. These models commonly replicate the 2D feature map on the depth axis because they assume that X-rays are generated by orthogonal projection (\cref{fig:projection}). However, since X-rays are generated by the perspective projection, the orthogonal projection places the features of X-rays in the wrong position. To provide accurate features to the model to reconstruct CT, we place the 2D X-ray feature in a 3D space using the perspective projection method, which is how X-ray images are generated. Recent studies~\cite{ge2022x, kasten2020end, ratul2021ccx} utilize segmentation maps to improve reconstruction quality. A significant drawback of these approaches is the necessity of a segmentation map, which is laborious to obtain.  
In contrast, our model reconstructs the 3D CT volume using only two X-rays (PA and lateral view images) without additional information.
We utilize two X-rays as input since a single X-ray image has insufficient depth information for accurate reconstruction of a 3D volume~\cite{shen19patrecon, ying2019x2ct}.

\begin{figure*}[hbt!]
\begin{center}
\includegraphics[width=0.8\linewidth]{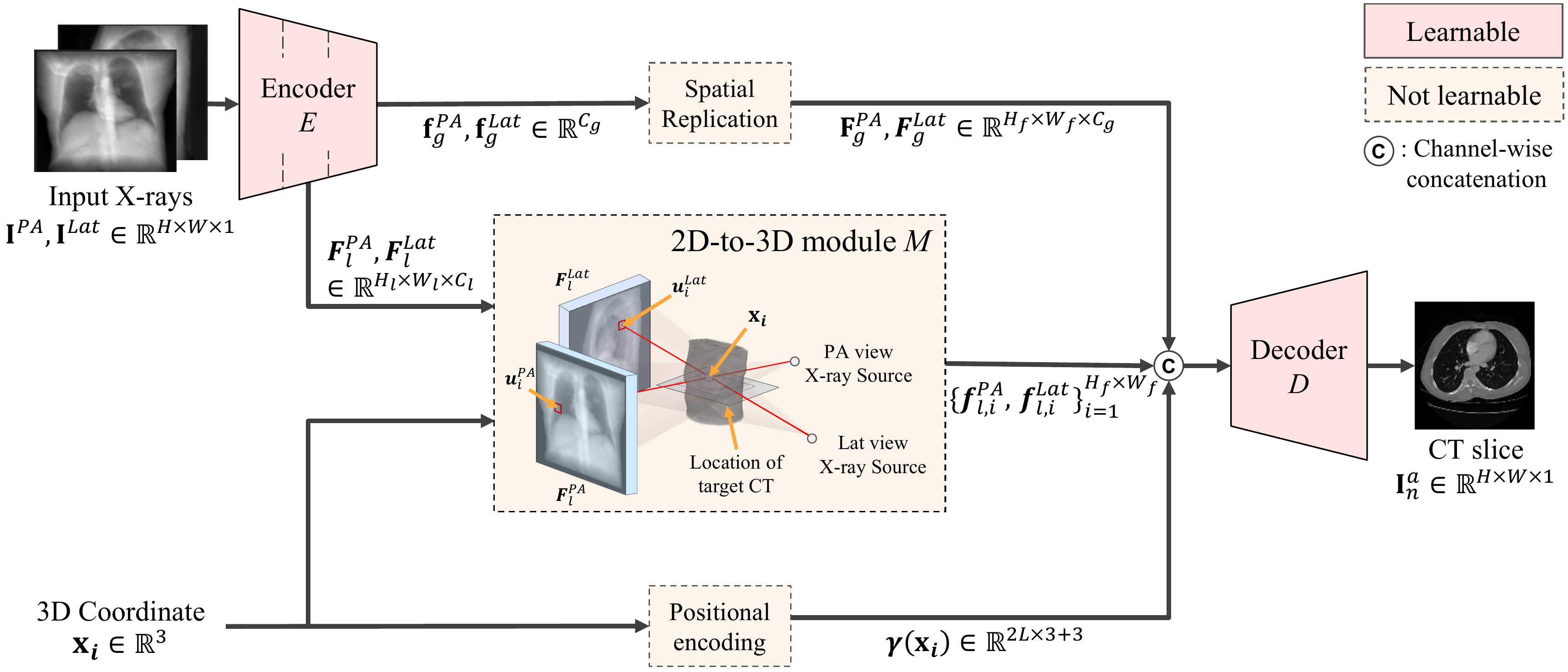}
\end{center}
\vspace{-0.4cm}
\caption{Overview of PerX2CT. The encoder extracts the local and global features of two X-rays and \textit{2D-to-3D module} extracts the local latent vector of target slice. The decoder reconstructs the output patch using the corresponding local, global feature map and position encoding.
}
\vspace{-0.4cm}
\label{fig:method overall}
\end{figure*}

Another limitation of previous studies is that they require high computational cost due to the heavy 3D decoder. Since the orthogonal projection places the same features across the entire depth axis, the model receives deficient information.
To compensate this shortage, the previous models reconstruct CT volume at the same time using the 3D decoder, incurring significant computational cost. However, we show that the 2D decoder is sufficient to reconstruct accurate CTs than the 3D decoders with the perspective projection, better suited for the X-ray image.
With a 2D decoder, the time complexity will be significantly reduced and each CT slice can be individually reconstructed efficiently. Nonetheless, the 2D decoder may lack information on the whole body compared to the 3D decoder. We thus additionally provide global features of the X-rays and the position of each voxel. 

In this work, we propose a simple yet efficient CT reconstruction framework that reflects the X-ray perspective projection scheme, named PerX2CT. The proposed method significantly improves the visual quality of reconstructed CT by accurately placing the information acquired from 2D X-rays in the 3D space. With coordinate-wise features in the 3D space and the 2D decoder, we achieve not only a reduced model complexity but also a 10x faster inference time. In addition, it can perform partial reconstruction for arbitrary selected patches with high resolution thanks to its ability to utilize 3D positional information. 

\section{Methodology}
In this part, we present our proposed CT reconstruction framework from X-rays taken from two perpendicular views, PA and lateral (\cref{fig:method overall}). We extract image features from X-rays and place them through the perspective projection method. Then, we reconstruct the target CT slice properly utilizing both local and global features.

\subsection{Feature Extraction from X-rays}
We extract image features from two X-rays with an encoder $E_{\theta}$ and place them in a 3D space to generate feature maps according to the target CT slice. 
To maximize the richness of information while minimizing the number of input X-ray images, we input two images of perpendicular views $\mathbf{I}^{PA}$, $\mathbf{I}^{Lat}\in\mathbb{R}^{H\times W\times 1}$ to $E_{\theta}$, where $H\times W$ denote the spatial resolution of X-ray images. For simplicity, we denote the view of an image as $v \in \{PA, Lat\}$.

We encode local and global features for each view to represent the voxel-specified information and whole-body information, respectively. Since each pixel value of the X-ray is independently calculated using voxels through which each X-ray beam passes, appropriate allocation of local feature using the path of each ray helps reconstructing each voxel of CT.
However, since local features only provide position-specified information, additional global features of X-rays help the model understand the whole body of a patient.
Thus, we extract low-level local feature maps $\mathbf{F}^{v}_{l} \in\mathbb{R}^{H_{l}\times W_{l}\times C_{l}}$ and high-level global feature vectors $\mathbf{f}^{v}_{g} \in\mathbb{R}^{C_{g}}$ from the encoder and provide them to the decoder as input. $C_l$ and $C_g$ are the channel size of each feature and ($H_l \times W_l$) denotes the spatial dimension of the local feature map.
\begin{equation}
    E_{\theta}({\mathbf I}^{v}) = (\mathbf{F}^{v}_{l}, \mathbf{f}^{v}_{g}),
\end{equation}

\begin{table*}[t!]
  \begin{center}
    {\small{
    \resizebox{0.9\linewidth}{!}{%
        \begin{tabular}{lcccccc}
            \toprule
               & PSNR(\(\uparrow\)) & SSIM(\(\uparrow\)) & LPIPS(\(\downarrow\)) & Params (M) & FLOPs (T) & Inference time (ms)\\ 
            \midrule            
            2DCNN~\cite{henzler17singlexray}& $25.398\pm0.038$   & $0.641\pm0.001$ & $0.483\pm0.001$ & 9.068 & 0.523 & $35.49\pm1.73$ \\ 
            X2CT-GAN-B~\cite{ying2019x2ct}    & $26.013\pm0.004$ & $0.666\pm0.003$ & $0.349\pm0.002$ & 72.796 & 1.207 & $1560.87\pm7.72$ \\ 
            \midrule
            PerX2CT                           & $\mathbf{27.450\pm0.016}$ & $\mathbf{0.732\pm0.000}$ & $\mathbf{0.213\pm0.001}$ & 42.536 & 0.178 & $46.643\pm1.43$  \\ 
            $\text{PerX2CT}_{global}$         & $\underline{27.335\pm0.004}$ & $\underline{0.725\pm0.001}$ & $\underline{0.216\pm0.003}$ & 70.131 & 0.190 & $52.82\pm1.74$  \\ 
            \bottomrule
        \end{tabular}
}}}
\end{center}
\vspace{-0.5cm}
\caption{Quantitative comparisons of CT reconstruction on the test set (mean ± std). Best and second best results are in \textbf{bold} and $\underline{\text{underlined}}$. 
}
\vspace{-0.5cm}
\label{table:test_all}
\end{table*}

\subsection{Resampling Local Feature via the 2D-to-3D Module}
The main role of the 2D-to-3D module $M$ is relocating the coordinate-wise feature vector $\mathbf{f}_{l,i}^{v}$ extracted from X-ray local feature map $\mathbf{F}_l^{v}$ to a 3D CT space (\cref{fig:method overall} (middle)). $i$  is an index for 3D coordinates $\mathbf{x}_i$ on target CT slice, and detailed description is provided below.
\begin{equation}
    \label{eq:2dto3dmodule}
    \mathbf{f}_{l, i}^{v}
    = M(\mathbf{F}_{l}^{v},\mathbf{x}_{i}).
\end{equation}

This process is composed of two steps: 1) projecting the 3D coordinate $\mathbf{x}_i$ of a voxel into the corresponding 2D projection point $\mathbf{u}_i^v$ of each X-ray image plane, and 2) resampling the local feature vector $\mathbf{f}_{l, i}^{v}$ from local feature maps $\mathbf{F}_l^{v}$ of each X-ray image using the projection point $\mathbf{u}_i^v$. 

Let $\{\mathbf{x}_i\}_{i=1}^{H_f \times W_f}\in \mathbb{R}^{3}$ be grid sampling points on the target CT slice $\mathbf{I}_n^a$, where $n$ denotes the slice number of target CT, $a$ is the imaging plane of CT ($a \in \{\text{axial}, \text{coronal}, \text{sagittal}\}$), and $H_f$, $W_f$ the resolution of the resampled feature map.

For each point $\mathbf{x}_i$, we compute the corresponding projection point $\mathbf{u}_i^v \in \mathbb{R}^{2}$ in the image coordinates as follows:
\begin{equation}
    \label{eq:transform matrix}
    \mathbf{u}_{i}^{v} =  \left[\mathbf{R}^v(\mathbf{\theta})|\mathbf{t}^v \right] 
    \mathbf{x}_i,
\end{equation}
where $\mathbf{R}^v(\mathbf{\theta}) \in \mathbb{R}^{3 \times 3}$ is the rotation matrix and $\mathbf{t}^v \in \mathbb{R}^{3}$ is the translation vector, which are the extrinsic parameters of the X-ray source.

After that, we extract the feature vector $\mathbf{f}_{l, i}^{v}$ corresponding to the projection point $\mathbf{u}_i^v$ using bilinear interpolation:
\begin{equation}
    \label{eq:2dto3d_projection}
    \mathbf{f}_{l, i}^{v}
    = \mathbf{F}_{l}^{v}(\mathbf{u}_{i}^{v}).
\end{equation}
The entire feature map is obtained by calculating $\mathbf{f}_{l, i}^{v}$ for all $\{\mathbf{x}_i\}_{i=1}^{H_{f}\times W_f}$ and is provided to the decoder as input. 

\subsection{Decoding CT Slices}
We independently reconstruct each CT slice using a 2D decoder instead of a 3D decoder for model efficiency. Our decoder takes five inputs: local feature maps $\{\mathbf{f}_{l, i}^{PA}\}_{i=1}^{H_{f}\times W_f}$, $\{\mathbf{f}_{l, i}^{Lat}\}_{i=1}^{H_{f}\times W_f}$, global feature maps $\mathbf{F}_{g}^{PA}, \mathbf{F}_{g}^{Lat}$, and the positional encoding of $\{\mathbf{x}_i\}_{i=1}^{H_{f}\times W_f}$. As global feature vectors $\mathbf{f}_{g}^{v}$ have a lower resolution than local feature maps, global feature vectors are spatially replicated to have the same spatial resolution as local feature maps by
\begin{equation}
    \label{eq:global_replication}
    \mathbf{f}_{g}^{v} \in \mathbb{R}^{C_g} \rightarrow \mathbf{F}_{g}^{v} \in \mathbb{R}^{H_f \times W_f \times C_g}.
\end{equation}

We extract both local and global features on the two views and concatenate all of them channel-wise. 
The decoder $\mathbf{D_{\phi}}$ reconstructs the target slice $\mathbf{I}_{n}^a$ by
\begin{equation}
    \label{eq:decoder}
    \begin{split}
        \hat{\mathbf{I}}_{n}^a &= \mathbf{D_{\phi}}(\mathbf{F}_{g}^{PA}, \mathbf{F}_{g}^{Lat}, \{\mathbf{f}_{l, i}^{PA}, \mathbf{f}_{l, i}^{Lat}, \gamma(\mathbf{x}_i)\}_{i=1}^{H_{f}\times W_f}),
    \end{split}
\end{equation}
where $\gamma(\cdot)$ is the sinusoidal positional encoding~\cite{mildenhall2020nerf}, which maps each coordinate from $\mathbb{R}$ to $\mathbb{R}^{2L+3}$. To aggregate the global context, we use an attention layer for the lowest resolution feature map of the decoder. 

\begin{table}[t!]
  \begin{center}
    {\normalsize{
    \resizebox{1\linewidth}{!}{%
        \begin{tabular}{cccc|ccc}
            \toprule
            Projection & PE & Global & Decoder & PSNR (\(\uparrow\)) & SSIM (\(\uparrow\)) & LPIPS (\(\downarrow\)) \\ 
            \midrule
            orthogonal  & \checkmark &              & 2D                      & $23.542\pm0.118$ & $0.576\pm0.003$ & $0.292\pm0.003$ \\ 
            \midrule
            perspective & \checkmark &              & 2D                      & $\textbf{27.751}\pm\textbf{0.076}$ & $\textbf{0.749}\pm\textbf{0.003}$ & $\textbf{0.200}\pm\textbf{0.002}$ \\
            perspective & \checkmark & \checkmark   & 2D                      & $\underline{27.706\pm0.069}$ & $0.747\pm0.003$ & $\underline{0.202\pm0.003}$ \\
            perspective &            &              & 2D                      & $27.685\pm0.116$ & $\underline{0.748\pm0.005}$ & $\textbf{0.200}\pm\textbf{0.003}$ \\
            \midrule
            perspective &            &              & $\text{2D}_{mini}$ & $27.198\pm0.175$ & $0.727\pm0.012$ & $0.213\pm0.008$ \\
            perspective &            &              & $\text{3D}_{mini}$ & $26.294\pm0.196$ & $0.699\pm0.007$ & $0.302\pm0.005$ \\
            \bottomrule
        \end{tabular}
}}}
\end{center}
\vspace{-0.5cm}
\caption{Ablation study of PerX2CT on the validation set (mean ± std). Best and second best results are in $\textbf{bold}$ and $\underline{\text{underlined}}$.
}
\vspace{-0.3cm}
\label{tab:ours_div1}
\end{table}

\begin{table}[t!]
\renewcommand{\arraystretch}{1.5}
\renewcommand{\tabcolsep}{0.9mm}
    \begin{center}
    {\normalsize{
    \resizebox{1\linewidth}{!}{%
    \begin{tabular}{cc|ccc|ccc}
    \toprule
        &        & \multicolumn{3}{c|}{$64 \times 64$} & \multicolumn{3}{c}{$32 \times 32$} \\ \cline{3-8} 
     PE & Global & PSNR ($\uparrow$) & SSIM ($\uparrow$) & LPIPS ($\downarrow$) & PSNR ($\uparrow$) & SSIM ($\uparrow$) & LPIPS ($\downarrow$) \\
     \midrule
                &            & 25.563 ± 0.557 & 0.793 ± 0.030 & 0.325 ± 0.020 & 23.969 ± 0.499 & 0.735 ± 0.037 & 0.341 ± 0.013 \\
     \checkmark &            & \underline{26.527 ± 0.017} & \underline{0.841 ± 0.001} & \textbf{0.295 ± 0.001} & \underline{25.029 ± 0.052} & \underline{0.800 ± 0.003} & \underline{0.314 ± 0.001}  \\
     \checkmark & \checkmark & \textbf{26.645 ± 0.096} & \textbf{0.846 ± 0.004} & \underline{0.294 ± 0.003} & \textbf{25.701 ± 0.118} & \textbf{0.831 ± 0.004} & \textbf{0.307 ± 0.003} \\
     \bottomrule
    \end{tabular}
    }}}
    \end{center}
\vspace{-0.4cm}
\caption{Ablation study of PerX2CT on the validation set for partial reconstruction (mean ± std). Best and second best results are in \textbf{bold} and $\underline{\text{underlined}}$.
We evaluate our model for $64 \times 64$ and $32 \times 32$ resolution depending on the size of the cropping part.}
\vspace{-0.5cm}
\label{tab:div4_val}
\end{table}

\subsection{Partial Reconstruction}
Since the proposed method employs the coordinate-wise feature vector to reconstruct the CT slice, it can reconstruct the CT slice not only in full resolution (\ie, full-frame) but also in arbitrarily selected patches.
We added a cropped CT slice as training data to perform the partial reconstruction to train the model. Specifically, we randomly cropped the CT slice selected with the probability of $p_{part}$ during the training phase. 
The cropping resolution is selected from the range from ($H_{min}, W_{min}$) to ($H_{out}, W_{out}$), and the bilinear interpolation is used. $H_{min}, W_{min}$ are hyperparameters to determine the sampling range. If a cropped slice is given as a target, the 2D-to-3D module $M$ samples the coordinate-wise local feature $\mathbf{f}_{l, i}^v$ within that part. Thus, PerX2CT can reconstruct the part of the CT slice in detail by densely sampling the corresponding features.

\subsection{Overall Objective}
Our overall objective function is given by
\begin{equation}
    \label{eq:totalloss}
    \mathcal{L}_{tot} = \lambda_{rec}\mathcal{L}_{rec} + \lambda_{p}\mathcal{L}_{p},
\end{equation}
where $\mathcal{L}_{rec}$ is the reconstruction loss which is the pixel-wise mean square error (MSE) for each slice, $\mathcal{L}_{p}$ is the perceptual loss~\cite{zhang18lpips}. $\lambda_{rec}, \lambda_{p}$ control relative importance of the two losses. 

\begin{figure}[t]
  \vspace{-0.1cm}
  \begin{center}
    \includegraphics[width=0.8\linewidth]{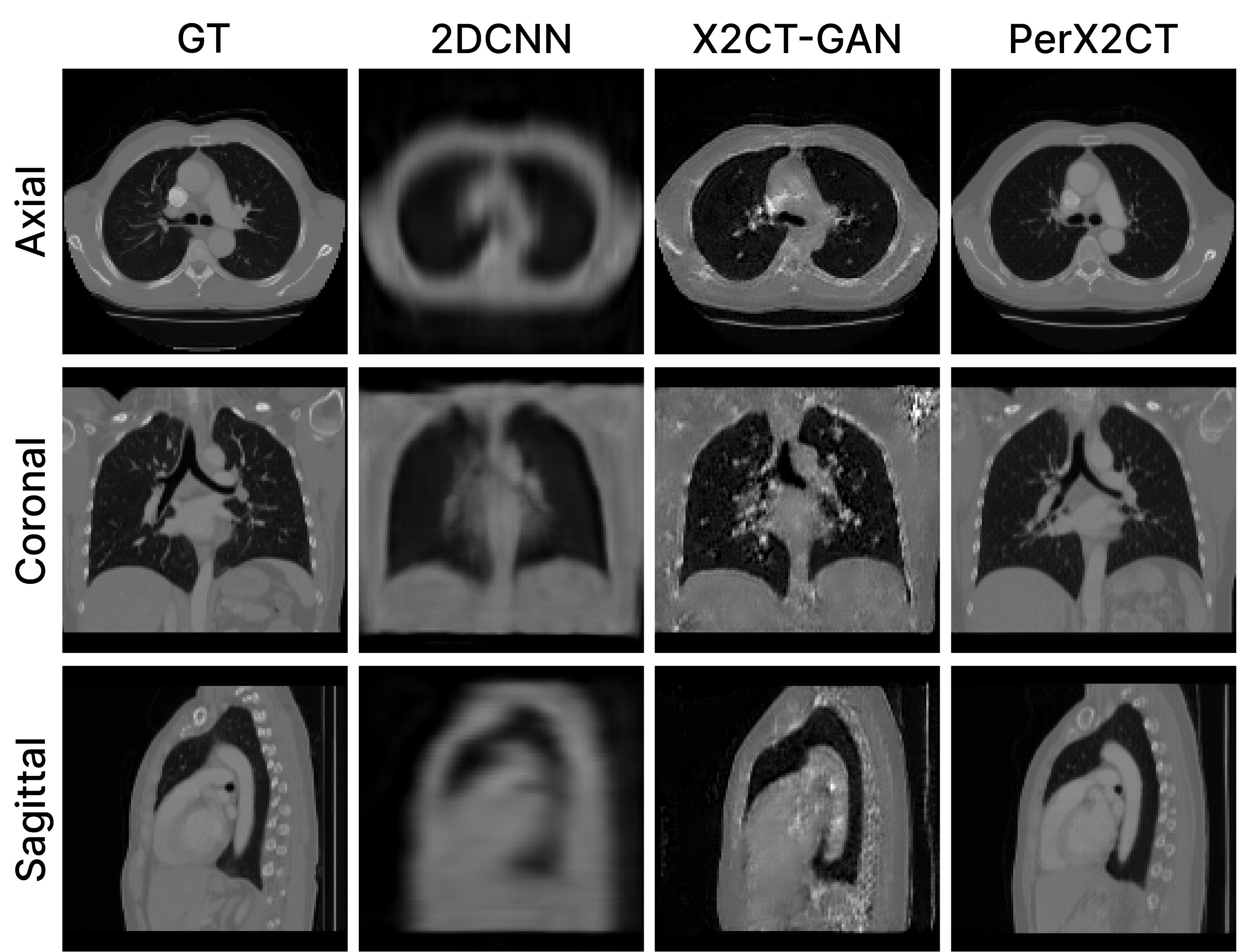}
  \end{center}
  \vspace{-0.45cm}
  \caption{Qualitative results of CT reconstruction. 
}
  \vspace{-0.55cm}
\label{fig:fullres}
\end{figure}

\section{Experiments}
\subsection{Experimental Settings} 
\textbf{Dataset.}
We require a dataset with X-ray and CT pairs to reconstruct CT from X-rays. Since collecting a real paired dataset is practically infeasible\footnote{We need an X-ray pair (PA and Lateral) and a CT scan that were taken simultaneously, which hardly happens in reality.}, existing studies have used digitally reconstructed radiograph (DRRs) technology to generate synthetic X-rays from a real CT volume~\cite{ying2019x2ct}. We construct the dataset following Ying \etal~\cite{ying2019x2ct}, using LIDC-IDRI~\cite{Armato11lidc}. We randomly split into 916, 20, and 82 for training, validation, and test set. For CTs, we use slice images for the three planes (\ie, axial, sagittal, and coronal) instead of full 3D data. 

\textbf{Evaluation Metrics.}
We use three different metrics to evaluate the quantitative quality of the predicted CT: peak signal-to-noise ratio (PSNR), structural similarity index (SSIM)~\cite{Wang04ssim}, and learned perceptual image patch similarity (LPIPS)~\cite{zhang18lpips}. Since PSNR and SSIM do not properly reflect perspective quality~\cite{zhang18lpips, lee16ssim, Wang04psnr}, we use LPIPSs~\cite{zhang18lpips} which focuses on latent semantic perception and is highly correlated with human visual perception.

\textbf{Implementation Details.}
For the image encoder, we fine-tune the ResNet-101\cite{He2016DeepRL} pre-trained on ImageNet as the backbone. The local feature map $\mathbf{F}_{l}^{\mathit{v}}$ and the global vector $\mathbf{f}_g^v$ are extracted from the third and fourth layer blocks, respectively.
The channel size of each feature $C_l$, $C_g$ are 512, 256, and the spatial dimension of the local feature map $H_l \times W_l$ is $16 \times 16$. Our proposed model resamples the local feature map $\{\mathbf{f}_{l, i}^{v}\}^{H_f \times W_f}_{i=1}$, with set $H_f \times W_f$ as $32 \times 32$ resolution from $\mathbf{F}_{l}^{v}$. Without global feature, we use the encoder up to the third-layer block, and all other settings are the same.
The decoder architecture follows~\cite{Esser20taming}, and set the number of layer blocks as 3. The output resolution of our decoder is 128 $\times$ 128. For positional encoding, we use $L$ as 10. The weights in Eq.~\eqref{eq:totalloss} are $\lambda_{rec} = \lambda_{p} = 1$.
Each model train for 50 epochs (70 epochs for the partial reconstruction) with a batch size of 40. We use Adam optimizer~\cite{adam} with an initial learning rate $4.5 \times 10^{-6}$. The momentum decay rate $\beta_1$ and the adaptive term decay rate $\beta_2$ are 0.5 and 0.9, respectively. We use Pytorch~\cite{pytorch} on the RTX 3090 Ti. All experiments are repeated three times and the mean and standard deviation are reported. The code will be available at \url{https://github.com/dek924/PerX2CT}

\subsection{Results}
We compare the quantitative and qualitative results of our models with two baselines~\cite{henzler17singlexray, ying2019x2ct}. Quantitative results are summarized in \cref{table:test_all}.
PerX2CT only uses the perspective projection method and positional encoding, while $\text{PerX2CT}_{global}$ additionally uses the global latent vector.
We show that the proposed methods have improvements over the baselines in the overall metrics, LPIPS, PSNR, and SSIM.
We also compare the computational complexity of the models. We analyze it based on the number of trainable parameters, the number of the floating-point operation, and the inference speed. PerX2CT shows a 6.8$\times$ improvement in FLOPs and a 33.5$\times$ faster inference time compared to X2CT-GAN. 2DCNN is the best in terms of the number of parameters and inference speed, but FLOPs are about 3 times larger than PerX2CT and the quality of the reconstruction is significantly worse. 

\begin{figure}[t!]
    \begin{center}
    \includegraphics[width=0.85\linewidth]{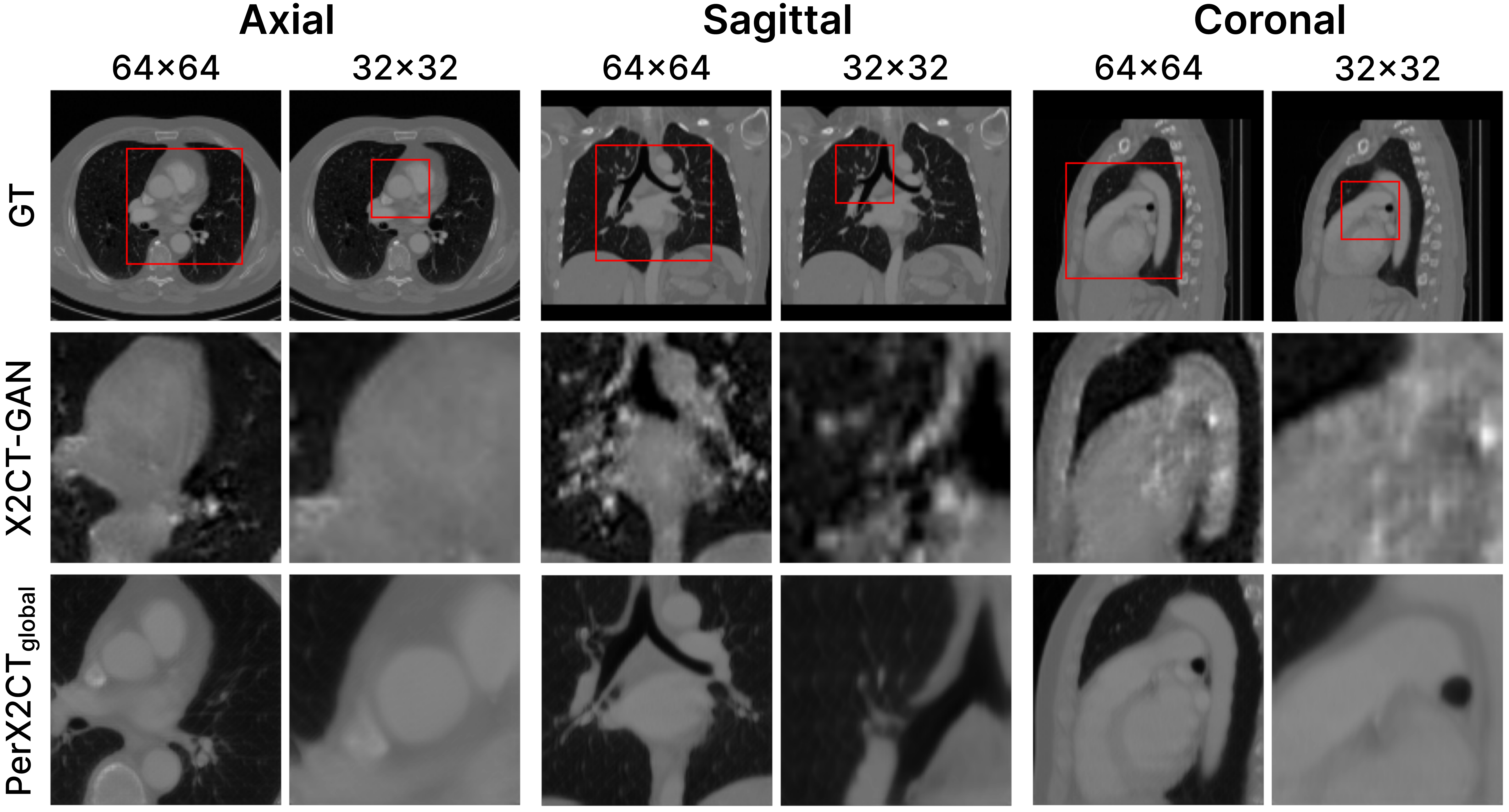}
    \end{center}
    \vspace{-0.4cm}
    \caption{Qualitative results for partial reconstruction. 
    The red bounding boxes denote the target patches for reconstruction, which have $64 \times 64$, and $32 \times 32$ resolutions.}
    \vspace{-0.6cm}
    \label{fig:div_arbitrary}
\end{figure}

In \cref{fig:fullres}, we compare the visual quality of our proposed model and the baselines. \cref{fig:fullres} shows that our reconstruction quality outperforms all baselines for all three views: axial, coronal and sagittal. 2DCNN reconstructs a blurrier image since it uses only one input X-ray. 
X2CT-GAN learns the boundary for large organs but fails to reconstruct accurately. PerX2CT successfully reconstructs the details of small anatomies, such as the atrium and aorta, clarifying the boundaries of the organs at the same time. 
\cref{fig:div_arbitrary} illustrates the visual quality of the partial reconstruction of $\text{PerX2CT}_{global}$ and the baselines. We show our results for the selected part on the $64 \times 64$ and $32 \times 32$ resolution. The baseline results are cropping from the full-frame slice reconstructed by each model and then up-scaling to a full-frame resolution through bilinear interpolation. Unlike this, $\text{PerX2CT}_{global}$ directly reconstructs the target part without interpolation. We simply achieve that by adding randomly cropped data at training without any model architecture change.

\subsection{Ablation Study}
We analyze the effectiveness of our model components by comparing the performance of the four variants in the validation set. 
In rows 1-2 of \cref{tab:ours_div1}, we evaluate the effectiveness of our feature expansion strategy, perspective projection, by showing a significant improvement for all metrics. We show the impact of positional encoding in both \cref{tab:ours_div1} and \cref{tab:div4_val}. PE also improves performance for all metrics. To clarify the benefit of our model architecture, we show the performance when all 2D convolutional layers in our model are converted to 3D. We conducted this experiment with a reduced number of layers, and without the attention layer due to memory issue for both 2D and 3D setting. As seen in rows 5-6 of the \cref{tab:ours_div1}, $\text{2D}_{mini}$ model outperforms the $\text{3D}_{mini}$ model. The global features have an insignificant effect on reconstructing the full-frame CT slice (rows 2-3 of \cref{tab:ours_div1}), but are effective for partial reconstruction (rows 2-3 in \cref{tab:div4_val}). The reason is that our decoder already aggregates the global context of the entire target CT slice through the self-attention layer. However, for partial reconstruction, the decoder receives only local features without additional global features.  

\begin{figure}[t!]
\begin{center}
\includegraphics[width=0.75\linewidth]{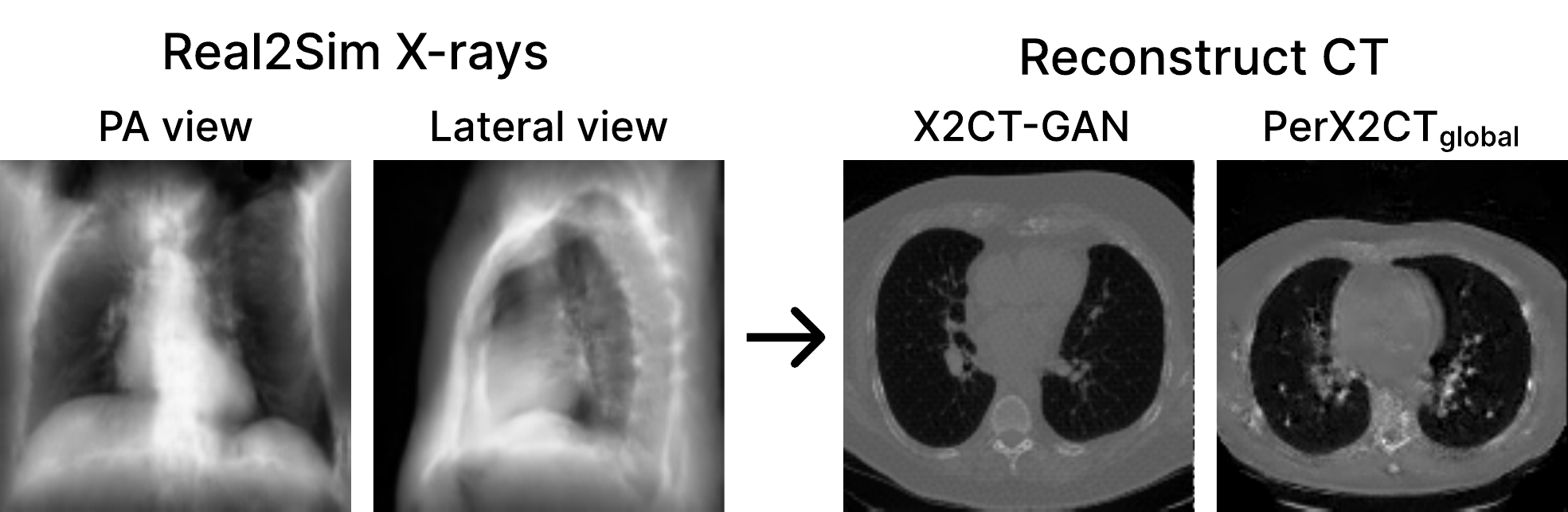}
\end{center}
\vspace{-0.4cm}
\caption{Qualitative results of the real X-ray projection.}
\vspace{-0.6cm}
\label{fig:real_sample}
\end{figure}

\subsection{Real-world Experiment}
To obtain an X-ray pair (PA and Lateral) and a CT scan that were taken exactly at the same time is hardly happens in reality. Therefore, we trained Real2Sim CycleGAN~\cite{CycleGAN2017} using 500 synthetic and real X-rays randomly selected from our training set and MIMIC-CXR~\cite{mimiccxrjpg}\footnote{The MIMIC-CXR-JPG data were available on the project website at https://www.physionet.org/content/mimic-cxr-jpg/2.0.0/}, respectively. After that, we translated real X-rays by CycleGAN and used them as input of our model. 
Because there is no ground-truth CT corresponding to the X-ray, it is impossible to evaluate accurately. Instead, we provide qualitative results in \cref{fig:real_sample}. As shown in \cref{fig:real_sample}, PerX2CT provides a clearer boundary than in X2CT-GAN, even in a real-world setting.

\section{Conclusion}
In this paper, we propose PerX2CT, a perspective projection-based 3D CT reconstruction framework. PerX2CT utilizes 3D coordinate-dependent local feature extracted from the 2D biplanar X-ray feature, reflecting the perspective projection.  We significantly improved the reconstruction performance of the full-frame slice while also reconstructing the desired part of the CT slice in a flexible manner. We validate that our proposed method outperforms existing models for all metrics. 
Our approach shows potential effectivenss in the medical field such as radiotherapy treatment planning with its high performance and being lightweight.
For future work, we plan to expand our dataset to reflect more realistic situations, such as the presence of metal on X-rays.

\bibliographystyle{IEEEbib}
\bibliography{strings,refs}

\end{document}